# Maximized Lateral Inhibition in Paired Magnetic Domain Wall Racetracks for Neuromorphic Computing

**Can Cui[1], Otitoaleke G. Akinola[1], Naimul Hassan[2], Christopher H. Bennett[3], Matthew J. Marinella[3], Joseph S. Friedman[2], and Jean Anne C. Incorvia[1]**

[1] Electrical and Computer Engineering, University of Texas at Austin, Austin, TX, USA
[2] Electrical and Computer Engineering, University of Texas at Dallas, Richardson, TX, USA
[3] Sandia National Laboratory, Albuquerque, NM, USA

Email: incorvia@austin.utexas.edu

**Abstract**

Lateral inhibition is an important functionality in neuromorphic computing, modeled after the biological neuron behavior that a firing neuron deactivates its neighbors belonging to the same layer and prevents them from firing. In most neuromorphic hardware platforms lateral inhibition is implemented by external circuitry, thereby decreasing the energy efficiency and increasing the area overhead of such systems. Recently, the domain wall – magnetic tunnel junction (DW-MTJ) artificial neuron is demonstrated in modeling to be inherently inhibitory. Without peripheral circuitry, lateral inhibition in DW-MTJ neurons results from magnetostatic interaction between neighboring neuron cells. However, the lateral inhibition mechanism in DW-MTJ neurons has not been studied thoroughly, leading to weak inhibition only in very closely-spaced devices. This work approaches these problems by modeling current- and field- driven DW motion in a pair of adjacent DW-MTJ neurons. We maximize the magnitude of lateral inhibition by tuning the magnetic interaction between the neurons. The results are explained by current-driven DW velocity characteristics in response to external magnetic field and quantified by an analytical model. Finally, the dependence of lateral inhibition strength on device parameters is investigated. This provides a guideline for the optimization of lateral inhibition implementation in DW-MTJ neurons. With strong lateral inhibition achieved, a path towards competitive learning algorithms such as the winner-take-all are made possible on such neuromorphic devices.

## 1. Introduction

Conventional von Neumann architecture has been the dominant large-scale computer architecture for the last five decades. Thanks to the rapid advancement of CMOS technology, shrinking transistor size and increased transistor density have been following Moore's law, e.g., each smaller node brings about both performance improvement and cost reduction. However, the throughput of a von Neumann computer is largely limited by the von Neumann memory wall [1], i.e. the separation of memory and central processing unit (CPU), and the sequential mode of instruction execution [2]; also, the von Neumann computer is energy-hungry due to the intensive data transfers between CPU and memory units [3]. In order to mitigate speed and power bottlenecks in the von Neumann architecture, research efforts have been directed towards the development of non-von Neumann computation paradigms with high parallelism and power-efficiency. The neuromorphic computing paradigm draws inspiration from the biological neural system, which consists of vast numbers of processing units, i.e. neurons, interconnected with synapses that carry the weights of neuron connectivity. Due to the in-memory computation nature and high parallelism, neuromorphic computing can outperform the von Neumann machine in speed and power efficiency [3-4].

The fundamental block of the artificial neural network (ANN) is the artificial neuron. It electrically mimics the biological neuron whose behavior can be described by an integrate-and-fire (IF) process [5]: the neuron receives electrical signals from its neighboring cells (reception), builds up its membrane potential (integration) and, once the potential exceeds a threshold voltage, generates a spike or action potential that is sent down to one or more post-synaptic cells (firing). The IF process omits many intricate biological details in favor of essential features of behavior, and is thus particularly useful in studying neural network dynamics. Extensions of the IF process include leaky integrate-and-fire (LIF) [6], adaptive quadratic integrate-and-fire [7], and adaptive exponential integrate-and-fire [8]. Some of these approaches have been adopted in neuromorphic computing platforms [9].

Lateral inhibition (LI) is another important neuron feature, closely associated with biological sensory systems. Receptive fields of tactile, auditory, and visual systems have center-surround responses to local stimuli: neurons pick up both presence of stimuli at the center and the absence thereof in the surrounding region, enhancing the signal contrast [5]. This function can only be achieved if central neurons inhibit the activity of peripheral, less-active neighbors in the same layer.

In neuromorphic computing, LI is crucial to the winner-take-all (WTA) algorithm [10-11]: in a neuron layer, mutual inhibition of the neurons should be strong such that only the most active neuron can produce a spiking output. The system's ability to pick a winner is necessary to competitive learning [12-14], pattern recognition [15-16], and general-purpose self-organizing networks [17]. It has also been shown to greatly improve the computing power of a neural network: for example, in one CMOS implementation of vector matrix multiplication, it was shown that including WTA gave a one-layer neural network the computing power equivalent to a two-layer neural network [18]. CMOS implementations of LI typically require additional circuit components such as differential amplifiers [19], a global reference voltage [20], or feedback loops [10]; in a hybrid memristor-MOS crossbar array [21], the inhibitory relation between neurons is realized by recurrently connecting neurons with memristor synapses. While LI or WTA functionalities have been successfully realized in these hardware platforms, the following drawbacks exist: 1) peripheral circuitry reduces power efficiency; 2) circuit design and layout are of great complexity; and 3) occupied chip area significantly increases with larger neuron numbers and connectivity. The overhead and energy cost is non-negligible in larger systems: for example, in one CMOS-based WTA implementation, 5 additional transistors are required per output neuron of a layer [22]. Therefore, an energy-efficient, simple, and scalable LI implementation is highly desirable.

Recently, a LIF neuron called domain wall – magnetic tunnel junction (DW-MTJ) neuron was demonstrated in simulation to be inherently inhibitory via magnetic interactions [23]. The neuron prototype is based on the three-terminal magnetic DW logic device [24] shown in the figure 1(a) side-view cartoon. It consists of a perpendicularly magnetized wire containing a single DW and an MTJ sitting on top of the wire. When current of density $J_e$ is applied to the wire, the DW propagates along the $+x$ direction through spin transfer torque (STT) or spin orbit torque (SOT). The MTJ defines the firing point of the neuron: when the DW moves past the junction, the wire magnetization under the MTJ is aligned with the top pinned ferromagnet layer, switching the MTJ resistance state low and generating a spiking current $I_{OUT}$ at the MTJ output terminal, which can be grounded at the subsequent device. Since DW velocity $v_{DW}$ increases with current density $J_e$, the neuron with higher current density has a higher chance to fire, and is therefore more active.

The inhibitory relation between a pair of DW-MTJ neurons is illustrated by figure 1(b). The two neurons are referred to as the neuron of interest Neuron I and its neighbor Neuron N, each with a single DW named $DW_I$ and $DW_N$ respectively. The DWs are driven by electrical current with density $J_{eI} < J_{eN}$, so that the DW velocity $v_{DWI} < v_{DWN}$ and the active Neuron N will be the first to fire. DW$_I$ falls behind DW$_N$ and is subjected to a stray field $\vec{B}_{\text{stray}}$ from Neuron N in the $-z$ direction; on the contrary, DW$_N$ experiences a stray field of Neuron I of same magnitude in the $+z$ direction. Thus, the magnetic force experienced by a DW is determined by its relative position with its neighbor. We will show that if the magnitude of the stray field in $-z$ direction is carefully chosen, it can serve as an inhibitory force to prevent the firing of the inactive neuron (Neuron I); on the contary, the stray field in $+z$ has much less impact on DW motion.

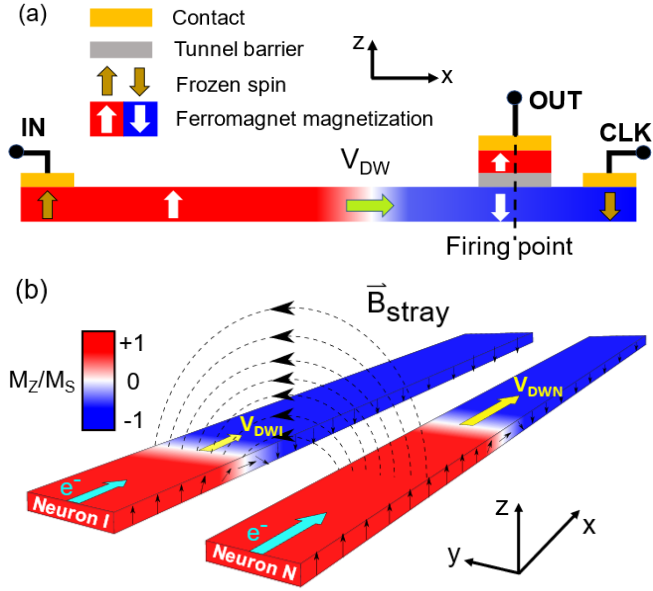

**Figure 1.** Domain wall – magnetic tunnel junction (DW-MTJ) neuron. (a) Cartoon of the structure of the DW-MTJ neuron, with colors and images defined by the legend. (b) Cartoon of a pair of adjacent DW-MTJ neurons (only the DW racetracks are shown). Here DW of Neuron I (left) is subjected to the stray field in $-z$ of the more active Neuron N (right), and can be inhibited. $M_s$: saturation magnetization; $M_z$: magnetization in $+z$.

This work focuses on investigating the LI mechanism, maximizing LI in a pair of DW-MTJ neurons, and understanding the design parameters to tune the LI based on both material and device parameters. Current- and field- driven DW motion is first simulated for the two-neuron system to find the optimal stray field magnitude for LI. The results are then explained by modeling the velocity characteristic of current-driven domain wall motion in a single neuron in response to an external magnetic field. We further quantify our simulation results with calculations based on the Landau-Lifshitz-Gilbert (LLG) equation, which reveal the impact of device geometry and material parameters on the magnitude of the largest achievable LI.

## 2. Methods

We model a pair of side-by-side magnetic wires with perpendicular magnetic anisotropy (PMA), each containing a single DW driven by electrical current via STT. We assume the MTJ output has negligible contribution to stray field and is omitted in the simulation. Both wires have dimensions 5 μm × 50 nm × 1.3 nm in length ($x$), width ($y$) and thickness ($z$), and are spaced $s$ nm apart in $\hat{y}$. The wire width is chosen to be large enough to investigate this effect for feasibly-fabricated prototypes; the results can be scaled to smaller widths and spacings. All simulations are carried out in Mumax3 [25]. Simulation cell size is 2 nm × 5 nm × 1.3 nm and material parameters are those of CoFeB [26]: saturation magnetization $M_s = $ 1273 emu cm$^{-3}$, anisotropy constant $K_U = 1 \times 10^7$ erg cm$^{-3}$, exchange stiffness $A_{\text{ex}} = 1.3 \times 10^{-6}$ erg cm$^{-1}$, Gilbert damping constant $\alpha = 0.02$, STT non-adiabatic parameter $\beta = 0.04$, and spin polarization $P = 0.72$. As above, whether a neuron can be inhibited depends on the magnitude of its activity relative to its neighbor's activity. In terms of the DW-MTJ neuron whose activity is encoded in DW velocity $v_{\text{DW}}$, LI can be quantified based on the reduction of $v_{\text{DW}}$ when a neuron is inhibited:

$$\text{LI} = \frac{v_{\text{DW (non-inhibition)}} - v_{\text{DW (inhibition)}}}{v_{\text{DW (non-inhibition)}}} \times 100\% \quad (1)$$

Denoting the DWs in the two wires as the DW of interest DW$_I$ and its neighbor DW$_N$, the two conditions of DW$_I$ motion are: (a) inhibition condition $J_{\text{eI}} < J_{\text{eN}}$ and (b) non-inhibition condition $J_{\text{eI}} > J_{\text{eN}}$. At simulation time $t = 0$ ns, a Néel-type DW is initialized at $x = 0$ nm for each wire to satisfy the fair start condition; electrical currents are then applied to both wires driving DW$_I$ and DW$_N$ along $+x$. For inhibition condition $J_{\text{eI}} = 2.2 \times 10^{12}$ A m$^{-2}$ and $J_{\text{eN}} = 4 \times 10^{12}$ A m$^{-2}$; for non-inhibition condition $J_{\text{eI}} = 2.2 \times 10^{12}$ A m$^{-2}$ and $J_{\text{eN}} = 0$ A m$^{-2}$. DW positions and velocities are extracted from the time evolution of the wire magnetization and LI is then calculated according to Eqn. (1).

## 3. Results

We first investigate the dependence of LI on the magnitude of magnetostatic interaction. For this purpose, we vary neuron spacing $s$ from 10 nm to 150 nm by steps of 10 nm and simulate the corresponding DW$_I$ velocity $v_{DWI}$. In figure 2(a), DW$_I$ position $x$ as a function of time $t$ for $s = 30$ nm, 60 nm, 90 nm, and 120 nm under inhibition and non-inhibition conditions are compared. It can be seen that as $s$ increases, the inhibited motion of DW$_I$ exhibits two distinct regions of behavior: in the strong magnetic interaction regime $s < 90$ nm, DW$_I$ has non-linear motion due to DW magnetization precession, i.e. Walker breakdown (WB) [27]: the DW$_I$ magnetization precesses in the $xy$ plane as it translates along the wire. In the weak magnetic interaction regime $s > 90$ nm, DW$_I$ has linear motion after a short settling time. Here, the weaker field from the neighbor brings the DW$_I$ magnetization orientation in the $xy$ plane (i.e. DW angle) to a fixed angle. In contrast, under the non-inhibition condition DW$_I$ has precessional motion (above WB) for every wire spacing simulated, though with different precession frequencies. DW$_I$ velocities $v_{DWI}$ for the inhibition and non-inhibition cases and corresponding LI as a function of $s$ are shown in figure 2(b). $v_{DWI}$ is taken as an average value in case of precessional motion, and the settled constant velocity otherwise. For the chosen material and geometry conditions, at $s = 90$ nm we see that $v_{DWI}$ is drastically reduced from 79 m s$^{-1}$ under non-inhibtion condition to 20 m s$^{-1}$ under inhibition condition, and LI reaches a maximum of 75%. Based on neuron geometry, material and spacing $s = 90$ nm, we estimate the stray field acting on DW$_I$ in inhibition case to be $H_z = -9$ Oe [28]. Compared to the amount of LI shown in [23], here LI is largely maximized by means of optimizing wire interaction strength.

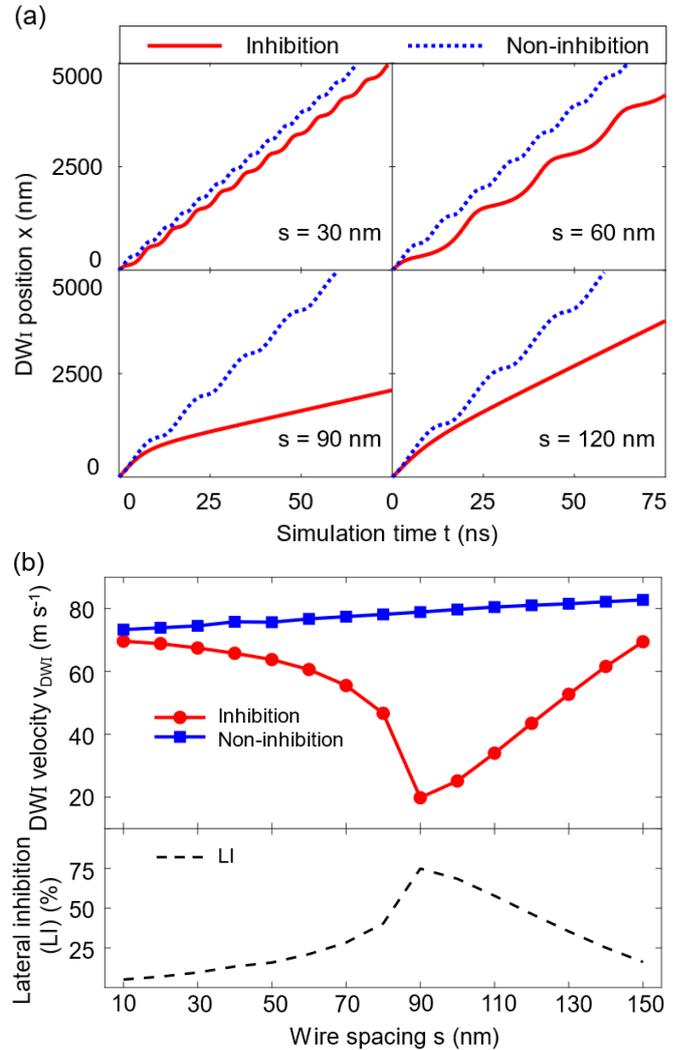

**Figure 2.** DW motion as a function of neuron spacing $s$. (a) DW$_I$ position $x$ as a funtion of time $t$ under inhibition (red solid) and non-inhibition (blue dotted) conditions for wire spacing $s = 30$ nm, 60 nm, 90 nm, and 120 nm; (b) (top) DW$_I$ velocity $v_{DWI}$ under inhibition (red circles) and non-inhibition (blue squares) conditions; (bottom) lateral inhibition LI as a function of $s$.

These results suggest that for a given neuron geometry, maximum LI is achieved when the magnetic interaction strength coincides with the WB field[4]. This is confirmed by approximating the influence of the neighboring neuron as a uniform vertical magnetic field $H_z$ and simulating the response of current-driven DW velocity of a single neuron to such field. Figure 3 shows the magnetic field that leads to WB when current density $J_e = 2.2 \times 10^{12}$ A m$^{-2}$.

---

[4] Strictly speaking, the Walker field includes the contribution from both current and magnetic field [29]. However, since we focus on the effect of magnetic field on DW motion and keep current density unchanged, we refer to Walker field as the magnetic field that leads to WB when current density $J_e = 2.2 \times 10^{12}$ A m$^{-2}$.

$v_{DW}$ as a function of $H_z$ ranging from −100 Oe to +100 Oe. For each data point, current density $J_e$ is held at 2.2 ×10$^{12}$ A m$^{−2}$ in simulation. A below-WB steady motion regime characterized by a high DW mobility $dv_{DW}/dH_z$ is observable. The regime is bounded by two Walker limits $H_{WL} = -9$ Oe $\pm 1$ Oe and $H_{WU} = -1$ Oe $\pm 1$ Oe, respectively, corresponding to lower-bound and upper-bound of $v_{DW}$. It is worth noting that even in the absence of an external magnetic field ($H_z = 0$ Oe) DW motion is already in WB regime; when $H_z < 0$ is applied, DW motion can be pushed back to the steady regime. Due to the high DW mobility in this regime, $v_{DW}$ can either be significantly increased (neuron excitation, $H_z = H_{WU}$) or decreased (neuron inhibition, $H_z = H_{WL}$). Thus, the maximum LI is achieved when the magnetic interaction strength is equal to $H_{WL}$, in good agreement with the optimal stray field of −9 Oe determined in the two-wire simulation.

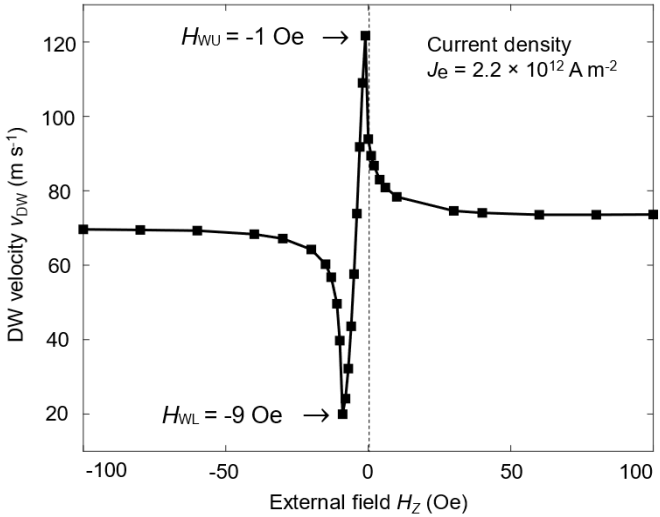

**Figure 3.** Current-driven DW velocity as a function of external magnetic field $H_z$. Current density is unchanged: $J_e = 2.2 \times 10^{12}$ A m$^{-2}$. Two Walker limits, $H_{WL} = -9$ Oe $\pm 1$ Oe and $H_{WU} = -1$ Oe $\pm 1$ Oe, are shown. As the guide to the eye, dotted line marks $H_z = 0$ Oe.

Having demonstrated the optimized magnetic interaction strength for LI given a set of device parameters, we next focus on maximizing LI in terms of geometric and material parameters. Time evolution of ferromagnet magnetization $\vec{M}$ is governed by the Landau-Lifshitz-Gilbert (LLG) equation. For the magnetic wire with a one-dimensional DW propagating in $x$, the LLG equation takes the form:

$$\frac{\partial \vec{M}}{\partial t} = \gamma \vec{H}_{\text{eff}} \times \vec{M} + \frac{\alpha}{M_s} \vec{M} \times \frac{\partial \vec{M}}{\partial t} - u\frac{\partial \vec{M}}{\partial x} + \beta u \frac{\vec{M}}{M_s} \times \frac{\partial \vec{M}}{\partial x} \quad (2)$$

Here $\gamma$ is gyromagnetic ratio, $\vec{H}_{\text{eff}}$ is the total effective magnetic field including external field $\vec{H}_{\text{ext}}$ and demagnetization field $\vec{H}_d$, and $u = (g\mu_B P/2eM_s)J_e$ is proportional to current density and has the dimensions of velocity, where $g$ is the Landé g-factor, $\mu_B$ is the Bohr magneton, and $e$ is the electron charge.

We use the macroscopic approach described in [29] which treats DW propagation as the result of different torques acting on DW; this gives the relation between the DW angle $\varphi$, vertical external field $H_z$, and $u$:

$$\sin 2\varphi = \frac{H_z + (\beta - \alpha)\frac{u}{\gamma\delta}}{2\pi\alpha M_s K_\perp}. \quad (3)$$

where $\delta$ is the DW width and hard axis anisotropy $K_\perp = N_x - N_y$ is the difference of DW demagnetization factors in $x$ and $y$ directions, and is proportional to the demagnetization energy difference between Neél- and Bloch-type DWs [30]. From Eqn. (3), $\varphi$ can have a time-independent solution $\varphi = \varphi_0$ only when the condition $|\sin 2\varphi| \leq 1$ is satisfied; otherwise $\varphi$ must be a time-varying quantity $\varphi = \varphi(t)$. We therefore obtain the two Walker limits $H_{WU}$ and $H_{WL}$:

$$H_{WU} = 2\pi\alpha M_s K_\perp - (\beta - \alpha)\frac{u}{\gamma\delta}, \quad (4)$$

$$H_{WL} = -2\pi\alpha M_s K_\perp - (\beta - \alpha)\frac{u}{\gamma\delta}; \quad (5)$$

and the conditions for steady and WB motions:

Steady,         $H_{WL} \leq H_z \leq H_{WU}$;

WB,             $H_z < H_{WL}$ or $H_z > H_{WU}$.

Accordingly, instantaneous $v_{DW}$ is a function of $\varphi_0$ or $\varphi(t)$:

Steady,    $v_{DW} = 2\pi\gamma\delta M_s K_\perp \sin 2\varphi_0 + u$,    (6)

WB,    $v_{DW}(t) = \frac{1}{1+\alpha^2} \times$
$[2\pi\gamma\delta M_s K_\perp \sin 2\varphi(t) + (1+\alpha\beta)u + \alpha\gamma\delta H_z].$  (7)

Given that $\alpha, \beta \ll 1$ and that the stray field magnitude is far smaller than the wire saturation field $H_z \ll 2\pi M_s$, Eqn. (7) takes the approximate form:

$$v_{DW}(t) \approx 2\pi\gamma\delta M_s K_\perp \sin 2\varphi(t) + u. \quad (8)$$

Comparing Eqns. (6) and (8) validates that given a weak stray field, instantaneous $v_{DW}$ has the same dependence on $\varphi$ for steady and WB motions. This is confirmed by extracting the $(\varphi, v_{DWI})$ relation from two-wire simulation results (figure 4). For $s = 60$ nm, DW$_I$ has WB motion and $\varphi$ changes in the range of $[0, 2\pi]$; as a result, $v_{DWI}$ changes with $\varphi$ and reaches the minimum $v_{min} = 20$ m s$^{-1}$ at $\varphi_{WL} = -\pi/4$ and the maximum $v_{max} = 125$ m s$^{-1}$ at $\varphi_{WU} = +\pi/4$. For the larger spacings $s = 90$ nm, 110 nm, 130 nm, and 150 nm, the stray field from the neighboring wire brings DW$_I$ to the steady motion regime, and $\varphi$ eventually settles to a fixed value. In such cases the $(\varphi, v_{DWI})$ relations are represented by single dots located on the $s = 60$ nm curve. Notably, $v_{min}$ at $\varphi_{WL} = -\pi/4$ is achieved for $s = 90$ nm. This confirms the drastic lowering of the $v_{DWI}$ at the optimized spacing earlier visible in figure 2(b), and therefore the large and maximized LI, arises from the neighboring wire's stray magnetic field setting $\varphi$ to the minimum velocity angle.

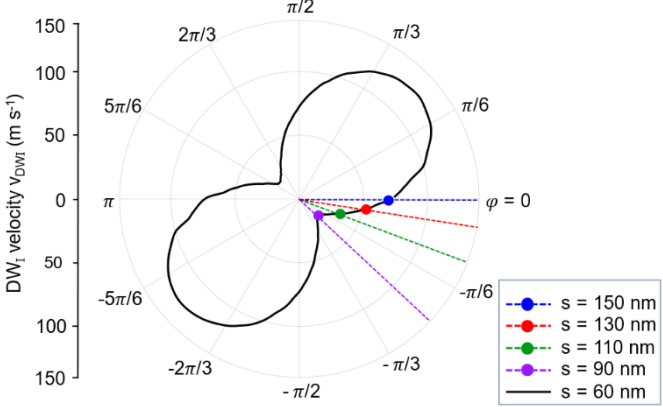

**Figure 4.** Dependence of instantaneous DW velocity $v_{DWI}$ on DW angle $\varphi$, extracted from the two-wire simulations. $v_{DWI}$ is plotted as spokes out from $v_{DWI} = 0$ m s$^{-1}$ at center, vs. $\varphi$. Both precessional ($s = 60$ nm) and steady ($s = 90$ nm, 110 nm, 130 nm and 150 nm) motions are shown.

Eqn. (6) can therefore be used to select the material and geometry parameters to maximize LI. Besides tuning the DW angle $\varphi$, the minimum velocity is equal to $-2\pi\gamma\delta M_s K_\perp + u$. Figure 5 summarizes the influence of saturation magnetization $M_s$, wire width $w$, and anisotropy constant $K_U$ on the largest achievable LI. For each set of parameters, a two-wire simulation is carried out and LI has been maximized in terms of wire spacing $s$.

Figure 5(a) shows that the LI is maximized for smaller wire widths $w$. We attribute the LI dependence on $w$ to the change of $K_\perp$. As is mentioned, $K_\perp$ is proportional to the demagnetization energy difference of Bloch and Neél walls. Bloch wall energy increases as $w$ becomes smaller because of the larger surface poles induced on the sidewalls, thereby increasing $K_\perp$. The impact of $M_s$ on LI is also shown in figure 5(a). Here, for each $M_s$ examined, we keep PMA quality factor $Q = K_U/2\pi M_s^2 = 1$ by choosing $K_U$ such that both $\delta$ and $K_\perp$ are mainly determined by wire aspect ratio $w/t$. For all $w$ the LI is largest for highest $M_s = 1273$ emu cm$^{-3}$. According to Eqn. (6), LI should be proportional to $M_s$; however, no substantial difference of LI is observed between $M_s = 1193$ erg cm$^{-3}$ and $M_s = 1114$ emu cm$^{-3}$: this is because although we keep $Q = 1$ to suppress the change of $\delta$, its slight increase for $M_s = 1114$ emu cm$^{-3}$ compared to $M_s = 1193$ emu cm$^{-3}$ is still sufficient to compensate for the reduction in $M_s$.

In figure 5(b), the saturation field is held at $M_s = 1273$ emu cm$^{-3}$ and LI is compared to the anisotropy constant $K_U$ for three wire widths: $w = 30$ nm, 40 nm and 50 nm. For each $w$, since hard axis anisotropy $K_\perp$ is independent of $K_U$, the decrease of LI with higher $K_U$ is mainly due to the shrinking of DW width $\delta$. Thus, by choosing small $w$ and keeping $Q$ close to 1, DW motion can be almost entirely halted by an inhibition of more than 90%, as is shown for $w = 30$ nm, $M_s = 1273$ emu cm$^{-3}$ and $Q = 1$.

(a)
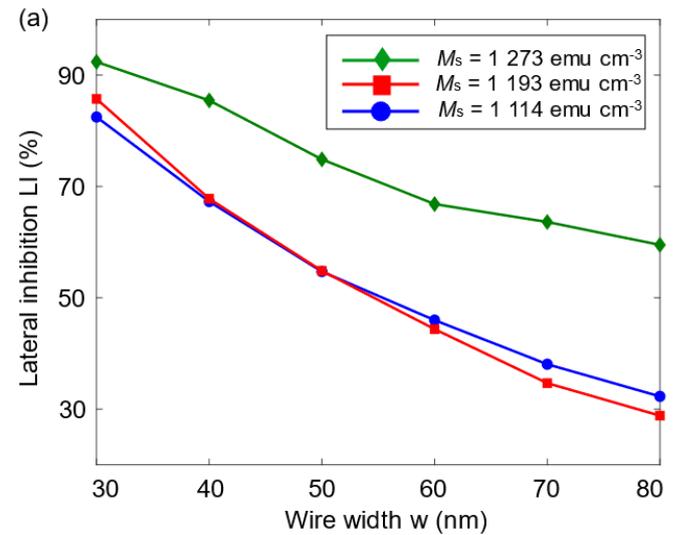

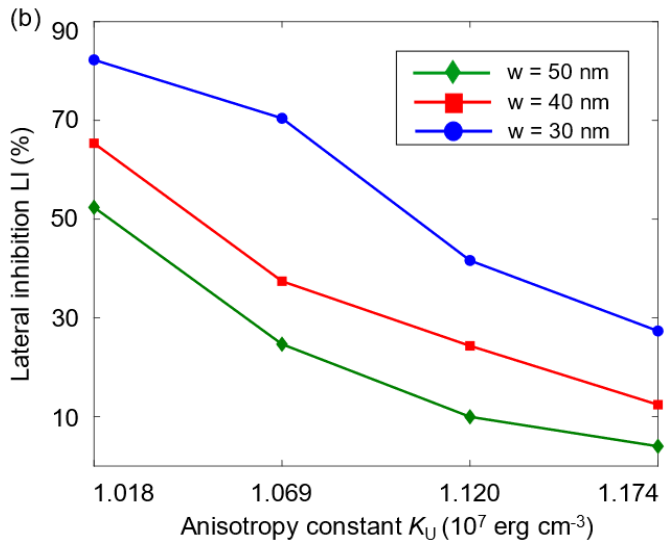

**Figure 5.** Dependence of maximum achievable lateral inhibition (LI) on wire width $w$, saturation magnetization $M_s$, and anisotropy constant $K_U$. (a) Maximized LI as a function of $w$ for $M_s = 1273$ emu cm$^{-3}$, 1193 emu cm$^{-3}$ and 1114 emu cm$^{-3}$; (b) maximized LI as a function of $K_U$, for $w = 30$ nm, 40 nm, and 50 nm.

## 4. Discussion

We here compare the magnitude of LI achieved in this work with that in [23]. In the previous work, an external magnetic field of −200 Oe is applied to implement leaking functionality. In such strong field the DW-MTJ neuron can only operate in the WB regime, wherein DW mobility is much lower compared to that of the linear regime. Therefore, adjacent neurons must be spaced close to achieve substantial LI. In order to implement the leaking feature in DW-MTJ neuron while maintaining a large LI, field-free implementation of leaking using shape or anisotropy gradients [31-32] can be adopted. It is worth noting that in these leaking implementations the DW can be already largely inhibited in the wire region close to the starting point, where wire width $w$ and aniotropy constant $K_U$ are small. Therefore, LI will not be degraded by the gradient-induced larger $w$ and $K_U$ close to the firing point.

## 5. Conclusions

An energy-efficient implementation of strong lateral inhibition in artificial neural networks is crucial to building competitive learning algorithms with emerging devices. This work proposes a method to maximize lateral inhibition in the domain wall – magnetic tunnel junction (DW-MTJ) neuron. By optimizing spacing between a pair of DW-MTJs, DW velocity is reduced by as large as 90% under inhibition condition (i.e. 90% lateral inhibition). Since this large inhibition does not require strong magnetostatic interaction strength in our implementation, adjacent DW-MTJs can be spaced further apart, enabling the fabrication of such devices with standard nanopatterning techniques. This work establishes a materially-feasible basis for inherent lateral inhibition in DW-MTJs, which can lead to future implementations of powerful neuro-inspired networks employing winner-take-all layers.

## Acknowledgements

The authors acknowledge discussions and funding from Sandia National Laboratories and computing resources from the Texas Advanced Computing Center (TACC) at the University of Texas at Austin (http://www.tacc.utexas.edu).

This paper describes objective technical results and analysis. Any subjective views or opinions that might be expressed in the paper do not necessarily represent the views of the U.S. Department of Energy or the United States Government. Sandia National Laboratories is a multimission laboratory managed and operated by NTESS, LLC, a wholly owned subsidiary of Honeywell International Inc., for the U.S. Department of Energy's National Nuclear Security Administration under contract DE-NA0003525.

## References

[1]  Backus J 1978 Can programming be liberated from the von neumann style? A functional style and its algebra of programs *Commun. ACM* **21** 613-641
[2]  Buehrer R and Ekanadham K 1987 Incorporating data flow ideas into von Neumann processors for parallel execution *IEEE Trans. Comput.* **100** 1515-1522
[3]  Merolla P A *et al.* 2014 A million spiking-neuron integrated circuit with a scalable communication network and interface *Science* **345** 668-673
[4]  Blouw P, Choo X, Hunsberger E and Eliasmith C 2019 Benchmarking keyword spotting efficiency on neuromorphic hardware *Neuro-inspired Computational Elements Workshop (NICE '19)*
[5]  Baars B J and Gage N M 2010 *Cognition, Brain, and Consciousness* (Cambridge: Academic Press) Chapter 3 Neurons and their connections
[6]  Stein R B 1965 A theoretical analysis of neuronal variability *Biophys. J.* **5** 173-194
[7]  Izhikevich E M 2003 Simple model of spiking neurons *IEEE Trans. Neural Netw.* **14** 1569-1572
[8]  Brette R and Gerstner W 2005 Adaptive exponential integrate-and-fire model as an effective description of neuronal activity *J Neurophysiol* **94** 3637-3642
[9]  Furber S 2016 Large-scale neuromorphic computing

systems *J. Neural Eng.* **13** 051001
[10] Fang Y, Cohen M A and Kincaid T G 1996 Dynamics of a winner-take-all neural network *Neural Networks* **9** 1141-1154
[11] Maass W 2000 On the computational power of winner-take-all *Neural Comput.* **12** 2519-2535
[12] Gupta A and Long L N 2009 Hebbian learning with winner take all for spiking neural networks *Proc. Int. Jt. Conf. Neural Networks* 1054-1060
[13] Ahalt S C, Krishnamurthy A K, Chen P and Melton D E 1990 Competitive learning algorithms for vector quantization *Neural Networks* **3** 277-290
[14] Urahama K, Nagao T 1995 K-winners-take-all circuit with O(N) complexity *IEEE Trans. Neural Netw.* **6** 776-778
[15] Gupta A and Long L N 2007 Character recognition using spiking neural networks *Proc. Int. Jt. Conf. Neural Networks* 53-58
[16] Cireşan D, Meier U and Schmidhuber J 2012 Multi-column deep neural networks for image Classification *arXiv preprint arXiv:1202.2745*
[17] Choi J, Sheu B J 1993 A high-precision VLSI winner-take-all circuit for self-organizing neural networks *IEEE J. Solid-State Circuits* **28** 576-584
[18] Ramakrishnan S and Hasler J 2013 Vector-matrix multiply and winner-take-all as an analog classifier *IEEE Trans. Very Large Scale Integr. Syst.* **22** 353-361
[19] Mead C A and Mahowald M A 1988 A silicon model of early visual processing *Neural Networks* **1** 91-97
[20] Lazzaro J, Ryckebusch S, Mahowald M A and Mead C A 1989 Winner-take-all networks of O(n) complexity *Adv. Neural Inf. Process Syst.* 703-711
[21] Ebong I E, Mazumder P 2011 CMOS and Memristor-based neural network design for position detection *Proceedings of the IEEE* **100** 2050-2060
[22] Mead C A, Allen T P, Synaptic Inc 1991 Adaptable CMOS winner-take all circuit *U.S. Patent 5,049,758*
[23] Hassan N, Hu X, Jiang-Wei L, Brigner W H, Akinola O G, Garcia-Sanchez F, Pasquale M, Bennett C H, Incorvia J A C and Friedman J S 2018 Magnetic domain wall neuron with lateral inhibition *J. Appl. Phys.* **124** 152127
[24] Currivan-Incorvia J A, Siddiqui S, Dutta S, Evarts E R, Zhang J, Bono D, Ross C A and Baldo M A 2016 Logic circuit prototypes for three-terminal magnetic tunnel junctions with mobile domain walls *Nat. Commun.* **7** 10275
[25] Vansteenkiste A, Leliaert J, Dvornik M, Helsen M, Garcia-Sanchez F and Van Waeyenberge B 2014 The design and verification of MuMax3 *AIP Adv.* **4** 107133
[26] Yamanouchi M, Jander A, Dhagat P, Ikeda S, Matsukura F, Ohno H 2011 Domain structure in CoFeB thin films with perpendicular magnetic anisotropy *IEEE Magn. Lett.* **2** 3000304
[27] Schryer N L and Walker L R 1974 The motion of 180° domain walls in uniform dc magnetic fields *J. Appl. Phys.* **45** 5406-5421
[28] Engel-Herbert R and Hesjedal T 2005 Calculation of the magnetic stray field of a uniaxial magnetic domain *J. Appl. Phys* **97** 074504
[29] Mougin A, Cormier M, Adam J P, Metaxas P J and Ferré J 2007 Domain wall mobility, stability and Walker breakdown in magnetic nanowires *EPL* **78** 57007
[30] Jung S W, Kim W, Lee T D, Lee K J, Lee H W 2008 Current-induced domain wall motion in a nanowire with perpendicular magnetic anisotropy *Appl. Phys. Lett.* **92** 202508
[31] Brigner W H, Hu X, Hassan N, Bennett C H, Incorvia J A C, Garcia-Sanchez F, Friedman J S 2019 Graded-anisotropy-induced magnetic domain wall drift for an artificial spintronic leaky integrate-and-fire neuron *IEEE J. Explor. Solid-State Computat.* **5** 19-24
[32] Brigner W H, Hassan N, Jiang-Wei L, Hu X, Saha D, Bennett C H, Marinella M J, Incorvia J A C, Garcia-Sanchez F, Friedman J S 2019 Shape-based magnetic domain wall drift for an artificial spintronic leaky integrate-and-fire neuron, *IEEE Trans. Electron Devices* **66** 4970-4975